\documentclass[12pt,a4paper]{article}

\usepackage[left=5em]{geometry}
\usepackage{amsmath}
\usepackage{graphicx}

\title{Naked Singularity in the Sultana-Dyer Space-Time}
\author{\textbf{Cheng-Yi Sun\footnote{cysun@mails.gucas.ac.cn; ddscy@163.com}\ $^{,a}$\
}\\ \\
 {$^a$\small Institute of Modern Physics, Northwest University,}\\
     \small Xian 710069, P.R. China.}

\begin{document}
\maketitle
\begin{abstract}
How to describe a black hole embedded in an expanding universe is an
important topic. Some models about this issue are suggested by
assuming that the metric is a conformal transformation of the
Schwarzschild metric or of the isotropic black hole metric. However,
there exists naked singularities in the two metrics. Recently, it is
argued that the singularity in the Sultana-Dyer space-time is
covered by an apparent horizon surface. But we find that such an
apparent horizon does not exist if the null energy condition holds.

\end{abstract}



\section{Introduction}
Isolated black holes have been investigated in great depth and
detail for more than forty years. On the other hand, black holes
embedded in the background of the expanding universe are also
important and even more realistic situations.  Some works on the
issue have been carried out. In \cite{McVittie}, McVitie found a
celebrated space-time describing a black hole embedded in the
Friedman-Robertson-Walker universe, which is generalized to the
Reissner-Nordstr\"{o}m case in \cite{g0407045}. In \cite{Swiss
cheese}, the model of Swiss cheese black holes  is shown. In
\cite{Vaidya}, the author suggested the Vaidya's space-time
describing a FRW universe with a Schwarzschild-like black hole that
does not expand with the rest of the universe, which is generalized
to the Kerr-Newman case in \cite{Patel}. In \cite{Thakurta}, the
Thakurta's black hole is shown. In \cite{Sultana and Dyer}, the
Sultana-Dyer black hole metric is suggested. Recently, in
\cite{McClure and Dyer,0707.1350}, the authors suggested new
solutions describing black holes embedded in the expanding universe.

Most of the models with the spherical symmetry on the topic are
based on the assumption
\begin{equation}
  \label{isotropicFRWBL}
  ds^2=-\left(\frac{1-\frac{m(t)}{2r}}{1+\frac{m(t)}{2r}}\right)^2dt^2+a^2(t)\left(1+\frac{m(t)}{2r}\right)^4(dr^2+r^2d\Omega^2),
\end{equation}
or
\begin{equation}
  \label{FRWSchwBL}
  ds^2=-\left(1-\frac{2m(t)}{r}\right)dt^2+a^2(t)\left(\frac{dr^2}{1-\frac{2m(t)}{r}}+r^2d\Omega^2\right),
\end{equation}
where $d\Omega^2=d\theta^2+\sin^2\theta d\varphi^2$. Here and after,
we take $c=G=\hbar=1$. However, the space-time manifold described by
the line element of Eq.(\ref{isotropicFRWBL}) is singular at
surfaces $r=m/2$. And there also exists a singularity in the
spacetime of Eq.(\ref{FRWSchwBL}) at the surface $r=2m$.

Then we may ask whether the singularity in the spacetime
(\ref{isotropicFRWBL}) or (\ref{FRWSchwBL}) is a spacetime
singularity or not? If it is a proper singularity, is the
singularity naked or covered by an apparent horizon? In
Ref.\cite{0802.1298,0907.4473}, the authors claimed that, in the
case of constant $m$, the singularity is covered by an apparent
horizon. However, we find such an apparent horizon does not exist if
the null energy condition is satisfied.

In the paper, firstly, we show that the space-time
(\ref{isotropicFRWBL}) or (\ref{FRWSchwBL}) is properly singular at
the surface $r=m(t)/2$ or $r=2m(t)$ respectively. Secondly, we show
the apparent horizon suggested in Ref.\cite{0802.1298,0907.4473}
does not exist. Section \ref{Conclusion} contains conclusions and
discussion.

\section{Proper singularity}
\label{proper}

The metric of the Mcvittie space-time \cite{McVittie} is assumed as
Eq.(\ref{isotropicFRWBL}) with $m(t)=\mu/a(t)$, where $\mu$ is a
constant parameter. In \cite{Sultana and Dyer,McClure and Dyer}, the
metric is assumed as Eq.(\ref{isotropicFRWBL}), but $m(t)$ is taken
to be constant. Here, we note that, if $\dot{m}\equiv dm/dt=0$, the
metrics in Eq.(\ref{FRWSchwBL}) and Eq.(\ref{isotropicFRWBL}) define
the same space-time. Generally, for $\dot{m}\neq0$, the two metrics
represent different spacetimes.  In Ref.\cite{0707.1350}, the
imperfect fluid stress-energy tensor is used to solve the Einstein
equations. In these models, at the surface $r=m(t)/2$, the spactimes
are singular. Let us show this by calculating the scalar curvature
of the metric in Eq.(\ref{isotropicFRWBL}),
\begin{equation}
  \label{isotropicSR}
  \begin{split}
    R_I=\frac{1}{B^3r}\left\{6A^2Br(\frac{\dot{a}^2}{a^2}+\frac{\ddot{a}}{a})
                             +3A(A+9B)\frac{\dot{a}}{a}\dot{m}+3(4B-A)\frac{\dot{m}^2}{r}+6AB\ddot{m}\right\}
  \end{split}
\end{equation}
where
\begin{equation}
  \label{AB}
  A=1+\frac{m(t)}{2r},\ \ \ B=1-\frac{m(t)}{2r},
\end{equation}
and $\dot{a}\equiv \frac{da}{dt}$ and $\ddot{a}\equiv
\frac{d^2a}{dt^2}$

Obviously, in the limiting cases corresponding to the standard FRW
universe, the standard Schwarzschild black hole or the
Schwarzchild-de Sitter black hole, the scalar curvature $R_I$ is
finite\cite{0908.3101}. However, in this paper, we exclude these
limiting cases. Then for $\dot{m}=0$, $R_I$ is infinity at the
surface $r=m/2$. Even for $\dot{m}\neq0$, we find that it is almost
impossible to find a time-dependent function $m(t)$ to avoid the
divergence at $r=m(t)/2$. The divergence of $R_I$ at $r=m/2$ implies
the singularity is proper.

The Thakurta spacetime \cite{Thakurta} is based on the line element
in Eq.(\ref{FRWSchwBL}). The scalar curvature of the metric
(\ref{FRWSchwBL}) is
\begin{equation}
  \label{SchwSC}
  R_S=\frac{2}{C}\left\{3(\frac{\dot{a}^2}{a^2}+\frac{\ddot{a}}{a})+4\frac{1}{C^2}\frac{\dot{m}^2}{r^2}
        +\frac{1}{C}(7\frac{\dot{a}}{a}\frac{\dot{m}}{r}+\frac{\ddot{m}}{r})\right\},
\end{equation}
where
\begin{equation}
  \label{C}
  C=1-\frac{2m(t)}{r}.
\end{equation}

Still, we exclude the limiting cases corresponding to the standard
FRW universe, the standard Schwarzschild black hole and the
Schwarzschild-de Sitter black hole. Then, similarly to the metric
(\ref{isotropicFRWBL}), the scalar curvature $R_S$ at the surface
$r=2m(t)$ is divergent. So the singularity at $r=2m$ is proper.

\section{Naked Singularity}
\label{naked}

In\cite{0802.1298,0907.4473}, the authors showed that, in the case
of $\dot{m}$=0, the singularity at the surface $r=m/2$ or $r=2m$ in
the space-time (\ref{isotropicFRWBL}) or (\ref{FRWSchwBL})
respectively is covered by an apparent horizon. In this section we
will show such an apparent horizon does not exist for the case of
$\dot{m}$=0 if the null energy condition holds.

First, we show the outgoing light emitted from the surface $r>m/2$
in Eq.(\ref{isotropicFRWBL}) or $r>2m$ in Eq.(\ref{FRWSchwBL}) can
reach the future null infinity asymptotically if $\dot{m}=0$.
Considering the world line of a photon propagating along the
direction of radius emitted from the point $(t=t_1,r=r_1)$ with
constant $\theta$ and $\phi$, we have
\begin{equation}
  \label{nullCurvesI}
  ds^2=0=-\left(\frac{B}{A}\right)^2dt^2+a^2(t)A^4dr^2,
\end{equation}
or
\begin{equation}
  \label{nullCurvesS}
  ds^2=0=-Cdt^2+\frac{a^2(t)}{C}dr^2,
\end{equation}
For an outgoing photon which arrives $r=r_2>r_1$ at the moment
$t=t_2>t_1$, we obtain respectively
\begin{equation}
  \label{timeAndRadiusI}
  \int^{t_2}_{t_1}{\frac{dt}{a}}=\int^{r_2}_{r_1}{\frac{(1+\frac{m}{2r})^3}{1-\frac{m}{2r}}dr}
\end{equation}
or
\begin{equation}
  \label{timeAndRadiusS}
  \int^{t_2}_{t_1}{\frac{dt}{a}}=\int^{r_2}_{r_1}{\frac{dr}{1-\frac{2m}{r}}}
\end{equation}
Of course, we assum $r_1>m/2$ in Eq.(\ref{timeAndRadiusI}) and
$r_1>2m$ in Eq.(\ref{timeAndRadiusS}), and $a(t)$ in the two
equations.

\begin{figure}
\centering
\renewcommand{\figurename}{Fig.}
\includegraphics[width=6.5cm]{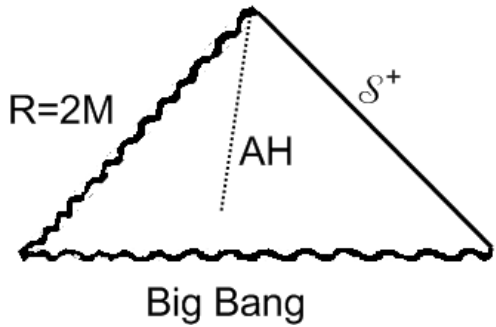}
\caption{\label{SultanaDyer}The conformal diagram of the
Sultana-Dyer spacetime. The horizontal wiggly line at the bottom
describes the Big Bang singularity, the wiggly line at 45 degrees
denotes the $R=2M$  null singularity, and the solid line at 45
degrees describes future null infinity. Null geodesics end at future
null infinity or at the black singularity (either when it is naked
if started early on, or crossing the timelike black hole apparent
horizon labelled AH).}
\end{figure}

Generally, since the Einstein equations which determine $a(t)$ and
$m(t)$ are unknown, nothing can be concluded from the equation
(\ref{timeAndRadiusI}) and (\ref{timeAndRadiusS}). But for the
special case of $\dot{m}=0$, the integral on the right-hand sides of
Eq.(\ref{timeAndRadiusI}) is finite for finite $r_2$. Then, in the
spacetime of Eq.(\ref{isotropicFRWBL}), the light emitted from the
point at the surface $r=r_1>m/2$ will reach the surface $r=r_2>r_1$
in the finite time. This implies that \emph{outgoing photons emitted
outside the surface $r=>m/2$ can reach the future null infinity
asymptotically in the spacetime (\ref{isotropicFRWBL}) if
$\dot{m}=0$}. In fact, this conclusion can be obtained easily from
the conformal diagram of the Sultana-Dyer space-time,
Eq.(\ref{isotropicFRWBL}) with $\dot{m}=0$, given in FIG.1 of
Ref.\cite{0907.4473} which is displayed as Fig.\ref{SultanaDyer} in
our paper (See Ref.\cite{0907.4473} for details).
For the space-time (\ref{FRWSchwBL}), similar conclusion can be made
from Eq.(\ref{timeAndRadiusS}).

Here, we note that, generally, to obtain our conclusion, we should
use the affine parameter $\tau$ in Eq.(\ref{timeAndRadiusI}) or
Eq.(\ref{timeAndRadiusS}) instead of the coordinate time $t$. But,
fortunately, here the coordinate time $t$ is appropriate for our
conclusion, since finite $t$ always indicates finite $\tau$ along
the worldline of an outgoing photon outside the surface $r=m/2$ or
$r=2m$ in the spacetime (\ref{isotropicFRWBL}) or (\ref{FRWSchwBL})
respectively.

Now let's assume that, outside the surface $r=m/2$ in the spacetime
(\ref{isotropicFRWBL}) or  the surface $r=2m$ in the spacetime
(\ref{FRWSchwBL}) respectively, there exists an apparent horizon
defined by the equation $f(t,r)=0$, if $\dot{m}=0$. Then at the
moment $t=t_0$, a space-like two surface is defined by $f(t_0,r)=0$,
and the expansion of the null generators of the future causal
boundary of the two surface, $\theta$, is zero at the two surface.
Due to the analysis in the last paragraph, we know outgoing photons
emitted from the surface $f(t_0,r)=0$ will reach the future null
infinity asymptotically if $\dot{m}=0$. Then, naturally, the
outgoing future causal boundary of the two surface intersects the
future null infinity.

However, on the other hand, it is known that the Raychaudhuri
equation \cite{Hawking,Wald,0809.3850} implies that the future
causal boundary of the two surface $f(t_0,r)=0$  cannot intersect
the future null infinity because $\theta$ is zero at the two surface
$f(t_0,r)=0$ and is nonincreasing if the null energy condition holds
\begin{equation}
  \label{NEC}T_{ab}l^al^b\geq0.
\end{equation}
$T_{ab}$ is the energy-momentum tensor and $l^a$ is a null vector.

Then we obtain two contrary conclusions from the assumption of the
existence of the apparent horizon. This indicates that \emph{in the
spacetime (\ref{isotropicFRWBL}) or (\ref{FRWSchwBL}), outside the
surface $r=m/2$ or $r=2m$ respectively, there does not exist any
apparent horizon and the singularity at $r=m/2$ or $r=2m$ is naked,
if the null energy condition holds and $\dot{m}=0$.}

In particular, in the model of the Sultana-Dyer black hole
\cite{Sultana and Dyer}, the energy-momentum tensor is
\begin{equation}
  \label{SDEMTensor}T_{ab}=T^{(I)}_{ab}+T^{(II)}_{ab}
\end{equation}
where $T^{(I)}_{ab} = \rho_I u_a u_b$ describes an ordinary massive
dust and $T^{(II)}_{ab} = \rho_n k_a k_b$ describes a null dust with
density $\rho_n$ and $k^a k_b=0$. Obviously, the null energy
condition is satisfied for any null vector $l^a$. Then we conclude
that in the space time of the Sultana-Dyer black hole, there is no
apparent horizon outside the surface $r=m/2$ in
Eq.(\ref{isotropicFRWBL}) and the singular surface $r=m/2$ is naked.
The conclusion is different to that in \cite{0907.4473}.

\section{Conclusion and Discussion}
\label{Conclusion}

The issue on the solutions of the Einstein equation describing black
holes embedded in the FRW universe is very important. Some spacetime
models on this issue \cite{McClure and Dyer,0707.1350,0810.2712}
have been suggested. Most of the models are based on the assumption
of Eq.(\ref{FRWSchwBL}) or Eq.(\ref{isotropicFRWBL}). However, there
exists singularity in the spacetime of Eq.(\ref{FRWSchwBL}) or
Eq.(\ref{isotropicFRWBL}). The singularity cannot be eliminated even
for the imperfect fluid stress-energy tensor to be used to solve the
Einstein equations \cite{0707.1350}. But it is argued that the
singularity in the spacetime (\ref{isotropicFRWBL}) at $r=m/2$ is
gravitationally weak \cite{Nolan} in the sense that it does not
crush extended bodies which falls into it to zero volume
\cite{Tipler}. In this view, the line elements
(\ref{isotropicFRWBL}) and (\ref{FRWSchwBL}) may still make some
sense in physics.

In \cite{0907.4473}, the authors claimed that, for $\dot{m}=0$,
there exists a regular apparent horizon $R_{bh}$ inside which the
surface $r=m/2$ in Eq.(\ref{isotropicFRWBL}) is contained. However,
our analysis above shows that photons emitted outside the surface
$r=m/2$ can reach the future null infinity if $\dot{m}=0$. So the
outgoing future causal boundary of the apparent horizon will
intersect the future null infinity. However, if the null energy
condition holds, it is impossible for the future causal boundary of
the apparent horizon to intersect the future null infinity because
of the non-decreasing expansion of the null generators of the causal
boundary indicates by the Raychaudhuri equation
\cite{Hawking,Wald,0809.3850}. Then such an apparent horizon does
not exist, and the singular surface is naked still if $\dot{m}=0$.
The conclusion in \cite{0802.1298,0907.4473} may be due to the
special coordinate $(T,R,\theta,\phi)$. Actually, the coordinate
transformation between $(T,R,\theta,\phi)$ and the ones in
Eq.(\ref{isotropicFRWBL}) is not diffeomorphic globally, which is
singular at the apparent horizon $R_{bh}$.

Our conclusion for the case of $\dot{m}=0$ cannot be applied to that
of $\dot{m}\neq0$ directly. However, since our analysis is based on
two conditions: the null energy condition holds and outgoing photons
emitted from the apparent horizon (if exists) can reach the future
null infinity. Then it is still possible for the existence of the
apparent horizon outside the surface $r=m/2$ or $r=2m$ in the
space-time (\ref{isotropicFRWBL}) or (\ref{FRWSchwBL}) respectively,
if the null energy condition is violated or outgoing photons emitted
from the apparent horizon (if exists) cannot reach the future null
infinity.

\section*{Acknowledgments}
This work is supported by the Natural Science Foundation of
Northwest University of China under Grant No. NS0927.

\end{document}